\documentclass[11pt]{article}
\usepackage{amsmath,amssymb}

\makeatletter
\@addtoreset{equation}{section}
\makeatother

\def\ben{\begin{equation}}
\def\een{\end{equation}}

    \let\e=\epsilon
    
  \let\n=\nu  \let\p=\pi

\let\C=\Chi

\def\nn{\nonumber} \def\bd{\begin{document}} \def\ed{\end{document}}
\def\ds{\documentstyle} \let\fr=\frac \let\bl=\bigl \let\br=\bigr
\let\Br=\Bigr \let\Bl=\Bigl
\let\bm=\bibitem
\let\na=\nabla
\let\pa=\partial \let\ov=\overline
\newcommand{\be}{\begin{equation}}
\newcommand{\ee}{\end{equation}}
\def\ba{\begin{array}}
\def\ea{\end{array}}
\def\ft#1#2{{\textstyle{\frac{\scriptstyle #1}{\scriptstyle #2} } }}
\def\fft#1#2{{\frac{#1}{#2}}}
\def\del{\partial}
\def\vp{\varphi}
\def\sst#1{{\scriptscriptstyle #1}}
\def\oneone{\rlap 1\mkern4mu{\rm l}}
\def\td{\tilde}
\def\wtd{\widetilde}
\def\ie{{\it i.e.\ }}
\def\dalemb#1#2{{\vbox{\hrule height .#2pt
        \hbox{\vrule width.#2pt height#1pt \kern#1pt
                \vrule width.#2pt}
        \hrule height.#2pt}}}
\def\square{\mathord{\dalemb{6.8}{7}\hbox{\hskip1pt}}}
\newcommand{\ho}[1]{$\, ^{#1}$}
\newcommand{\hoch}[1]{$\, ^{#1}$}
\newcommand{\bea}{\begin{eqnarray}}
\newcommand{\eea}{\end{eqnarray}}
\newcommand{\ra}{\rightarrow}
\newcommand{\lra}{\longrightarrow}
\newcommand{\Lra}{\Leftrightarrow}
\newcommand{\bp}{\tilde \beta^\prime}
\newcommand{\tr}{{\rm tr} }
\newcommand{\Tr}{{\rm Tr} }
\def\0{{\sst{(0)}}}
\def\1{{\sst{(1)}}}
\def\2{{\sst{(2)}}}
\def\3{{\sst{(3)}}}
\def\4{{\sst{(4)}}}
\def\5{{\sst{(5)}}}
\def\6{{\sst{(6)}}}
\def\7{{\sst{(7)}}}
\def\8{{\sst{(8)}}}
\def\n{{\sst{(n)}}}
\def\cA{{{\cal A}}}
\def\cB{{{\cal B}}}
\def\cF{{{\cal F}}}
\def\cG{{{\cal G}}}
\def\cH{{{\cal H}}}
\def\tV{\widetilde V}
\def\tW{\widetilde W}
\def\tH{\widetilde H}
\def\tE{\widetilde E}
\def\tF{\widetilde F}
\def\tA{\widetilde A}
\def\im{{{\rm i}}}
\def\tY{{{\wtd Y}}}
\def\ep{{\epsilon}}
\def\vep{{\varepsilon}}
\def\bD{{{\bar D}}}
\def\R{{{\mathbb R}}}
\def\C{{{\mathbb C}}}
\def\H{{{\mathbb H}}}
\def\CP{{{\mathbb C}{\mathbb P}}}
\def\RP{{{\mathbb R}{\mathbb P}}}
\def\Z{{{\mathbb Z}}}
\def\bA{{{\mathbb A}}}
\def\bB{{{\mathbb B}}}
\def\bC{{{\mathbb C}}}
\def\bD{{{\mathbb D}}}
\def\bE{{{\mathbb E}}}
\def\bZ{{{\mathbb Z}}}
\def\Re{{{\frak{Re}}}}
\def\Im{{{\frak{Im}}}}
\def\cosec{{\,\hbox{cosec}\,}}
\def\Gm{{\Gamma_{\!\! -}}}
\def\Gp{{\Gamma_{\!\! +}}}
\def\stan{{standard }}
\def\nonstan{{supernumerary }}
\def\p{{\partial}}
\def\kdel#1{{\fft{\del}{\del#1}}}

\def\bog{{Bogomolny }}
\def\om{{\omega}}

\newcommand{\tamphys}{\it George P. \& Cynthia Woods Mitchell  Institute
for Fundamental Physics and Astronomy,\\
Texas A\&M University, College Station, TX 77843, USA}

\newcommand{\damtp}{\it DAMTP, Centre for Mathematical Sciences,
 Cambridge University,\\  Wilberforce Road, Cambridge CB3 OWA, UK}

\newcommand{\auth}{
H. L\"u{$^1$}, Jianwei Mei{$^1$} and C.N. Pope{$^1,^2$}
}

\begin{document}

\begin{flushright}
\hfill{
MIFP-08-12}\\
\end{flushright}

\begin{center}

{\large {\bf New Charged Black Holes in Five Dimensions 
             
}}

\vspace{25pt}

\auth

\vspace{10pt}
{$^1$}{\tamphys}

\vspace{10pt}
{$^2$}{\damtp}

\vspace{25pt}

\underline{ABSTRACT}

\end{center}

     We obtain new stationary charged solutions of five-dimensional minimal
supergravity.  We first obtain purely dipole charged solutions, by
extending a technique that we developed for five-dimensional
Ricci-flat metrics in a previous paper, which could be viewed as being
analogous to a four-dimensional construction by Demianski and
Plebanski.  The further introduction of electric charge is achieved by
means of a solution-generating technique, which exploits the global
$SL(2,\R)$ symmetry of five-dimensional minimal supergravity reduced
on a timelike direction to four dimensions.  We present this procedure
in detail, since it provides a particularly simple general way of
adding charge to any stationary solution of five-dimensional minimal
supergravity.  The new charged solutions we obtain limit in special
cases to black rings carrying electric and magnetic dipole charge, or
to charged Myers-Perry rotating black holes.  We analyse the general
solutions in detail, showing that they can describe asymptotically
locally flat black holes whose horizon is a lens space
$L(n;m)=S^3/\Gamma(n;m)$.  At infinity they approach
Minkowski$_5/\Gamma(m;n)$.

\vspace{15pt}

\thispagestyle{empty}

\pagebreak
\setcounter{page}{1}

\tableofcontents

\addtocontents{toc}{\protect\setcounter{tocdepth}{2}}


\section{Introduction}

   String theory and M-theory provide a powerful motivation for expanding
the long-standing quest for solutions in four-dimensional general
relativity to higher dimensions.  As well as vacuum solutions of the
Einstein equations, solutions including the specific matter fields of
the relevant higher-dimensional supergravity theories are also of
interest.  With the recent discovery of the black ring solutions
\cite{empreal,ps} in five dimensions, some of the previous understanding
of black holes and their uniqueness, derived in a four-dimensional 
setting, needs to be extended.

In a previous paper \cite{lumeipope}, we looked for higher-dimensional
generalisations of the rather large class of four-dimensional
Ricci-flat type D metrics that are succinctly described in the form
found by Plebanski and Demianski \cite{plebdemi}.  Our starting point
was a reformulation \cite{chenlupope1,chenlupope2} of the
higher-dimensional rotating black holes, with
\cite{gilupapo1,gilupapo2} or without \cite{myper} cosmological
constant.  The reformulation in \cite{chenlupope1,chenlupope2} allowed
the natural introduction of additional parameters that could be
interpreted as certain higher-dimensional generalisations of the
four-dimensional NUT parameter.  The structure of the metrics in
\cite{chenlupope1,chenlupope2} so closely parallels the
four-dimensional Kerr-NUT-de Sitter metrics that one is tempted to
expect a parallel ``Plebanski-Demianski type" of generalisation in all
the higher dimensions too.  As was found in \cite{kk,lumeipope}, however,
this does not in general seem to be possible.  It was shown,
however, that five dimensions is rather a special case, and the
hoped-for generalisation was indeed found, for Ricci-flat metrics, in
that case \cite{lumeipope}.

   The purpose of the present paper is to extend the Ricci-flat results in 
\cite{lumeipope}, by constructing analogous solutions of five-dimensional
minimal supergravity.  This theory, in its bosonic sector (which is all
that is relevant for the solutions we shall consider), comprises 
five-dimensional gravity coupled to a $U(1)$ graviphoton field.  The
latter is effectively a Maxwell field in five dimensions, together with
a Chern-Simons type of $F\wedge F\wedge A$ coupling.  The new solutions
that we obtain carry both electric charge and also a magnetic dipole
charge.

     Our strategy for constructing the new solutions is to begin by
reformulating the previously known charged rotating black holes of
five-dimensional minimal supergravity (contained within the results in
\cite{cvetyoum} or in \cite{cclp}) in the manner of
\cite{chenlupope1}, and then to seek a generalisation analogous to the
one found for Ricci-flat metrics in \cite{lumeipope}.  Owing to the
fact that certain Wick rotations are performed in the process, the
interpretation of the new metrics at this stage is that they carry not
electric charge, but magnetic dipole charge.  We then, finally,
implement a solution-generating procedure that allows us to introduce
a genuine electric charge as well.  This solution-generating technique
exploits the $SL(2,\R)$ global symmetry of five-dimensional minimal
supergravity after a timelike reduction to four dimensions.  Since it is of
rather general utility, and is quite considerably simpler than other
presentations of the ``charging'' procedure that have appeared in the
literature, we devote a section of the paper to the description of
this procedure in general.  It can be applied to any stationary solution of
five-dimensional minimal supergravity.

Having obtained the new solutions with electric and magnetic dipole charge
we then
investigate their global properties.  The conditions on
the free parameters that must be imposed in order to eliminate conical
singularities imply that the spacetime describes a black hole with an
horizon that has the topology of a lens space $L(n;m)$ (a factoring of
$S^3$ by a certain freely-acting discrete subgroup $\Gamma(n;m)$ of
the isometry group of the sphere).  The metrics are asymptotically locally
flat, and at large distance they approach five-dimensional
Minkowski spacetime factored by $\Gamma(m;n)$.

     Finally, in appendix A, we present a more general class of
electrically-charged solutions to $U(1)^3$ five-dimensional
supergravity.  (This theory, sometimes called the STU model, 
can be thought of as a truncation of
maximal supergravity in five dimensions, in which two 
vector multiplets are coupled to the minimal supergravity.)  In
appendix B, we consider a certain limit of our general solutions in
minimal supergravity, in which the metrics have the
general form of black rings.  The requirement
that the solutions in this limit be globally well behaved 
leads to a bifurcation in the choice of the parameters.  In one
branch, this corresponds to turning off the electric charge.  This
solution, which still has a non-vanishing dipole charge, admits
choices for the remaining parameters for which it describes a black
ring with $S^2\times S^1$ horizon topology.  In the other branch, the
electric charge is non-zero but is related to the strength of the
dipole field.  Again, there exist black ring solutions within this
branch.  These black ring secialisations are contained within the results
found in \cite{emparan,elvempfig}.

\section{Charging Solutions of Minimal $D=5$ Supergravity}
\label{chargingsec}

   In this section, we present a general procedure for mapping any
stationary solution of five-dimensional minimal supergravity
into a solution that carries an additional electric charge.  Five-dimensional
minimal supergravity reduced on 2-torus gives rise to a scalar coset
sigma model with $G_2$ global symmetry \cite{mo,cjlp}.  One way to
charge a five-dimensional solution is to reduce it to $D=3$ on the
2-torus and then ``boost'' it by acting with an appropriate 
subgroup of the $G_2$ symmetry
\cite{bccgsw}.  However, it was observed in  \cite{cjlp} that reducing the
theory just on $S^1$ gives rise to a $D=4$ theory with an $SL(2,\R)$
global symmetry.  For our purposes, the charging can be 
conveniently implemented by performing a timelike reduction 
to four dimensions, and acting with a certain
$O(1,1)$ subgroup of the $SL(2,\R)$ duality group.  This procedure can be
carried out whenever the original five-dimensional solution is stationary.

   To describe the procedure, we first reduce five-dimensional minimal
supergravity on a timelike Killing vector.  In the bosonic sector, the
five-dimensional theory is given by
\be
{\cal L}_5 = \hat R\, {\hat*\oneone} - \ft12{\hat* \hat F}\wedge\hat F +
   \ft1{3\sqrt3}\, \hat F\wedge\hat F\wedge \hat A\,,\label{d5lag}
\ee
where $\hat F=d\hat A$, and $\hat A$ is the graviphoton potential.\footnote{
This implies that the graviphoton equation of motion is
$\hat\nabla_\nu \hat F^{\mu\nu} + 1/(4\sqrt3) \ep^{\mu \alpha\beta\gamma\delta}
  \hat F_{\alpha\beta} \hat F_{\gamma\delta}=0$.}  The
stationary metric can be written in the form
\be
d\hat s_5^2 = e^{\phi} ds_4^2  - e^{-2\phi}\, (dt + \cA)^2\,,\label{metans}
\ee
where the quantities on the right-hand side are all independent of $t$.

  We now perform a Kaluza-Klein reduction to four dimensions using
(\ref{metans}), with the reduction of the graviphoton given by
\be
\hat A = A + \sqrt3 \chi (dt+\cA)\,.\label{Aans}
\ee
The resulting four-dimensional theory is described by the Lagrangian
\be
{\cal L}_4 = R\, {*\oneone} -\ft32 {*d\phi}\wedge d\phi +
           \ft32 e^{2\phi} {*d\chi}\wedge d\chi +
       \ft12 e^{-3\phi}{*\cF}\wedge \cF - \ft12 e^{-\phi} {*F}\wedge F
  + \chi dA\wedge dA\,,\label{d4lag}
\ee
where
\be
\cF = d\cA\,,\qquad F= dA + \sqrt3 \chi d\cA\,.\label{pot1}
\ee

    The four-dimensional theory described by (\ref{d4lag}) has an
$SL(2,\R)$ global symmetry, at the level of the equations of motion.
By examining the dilaton coupling of the vector fields, it was
observed in \cite{cjlp} that the two vectors and their duals form
a quartet representation under the $SL(2,\R)$.  Here we shall exhibit
this structure explicitly.  A convenient way to do this is to use a
doubled formalism, in which potentials dual to $\cA$ and $A$ are
introduced. 

   First, we observe from (\ref{d4lag}) that the equations
of motion for $\cA$ and $A$ are given by
\be
d(e^{-3\phi} {*\cF}) -\sqrt3 e^{-\phi} d\chi\wedge {*F}=0\,,\qquad
d(e^{-\phi}{*F})- 2 d\chi\wedge dA -2\sqrt3 \chi d\chi\wedge d\cA=0
 \,.\label{Feoms}
\ee
We then define dual fields according to
\be
G = e^{-\phi} \, {*F}\,,\qquad \cG= e^{-3\phi}\, {*\cF}\,.
\label{duals}
\ee
It follows from (\ref{Feoms}) that we may write these in terms of dual
potentials $B$ and $\cB$ as
\bea
G&=& dB + 2\chi dA + \sqrt3 \chi^2 d\cA\,,\nn\\
\cG &=& d\cB + \sqrt3 \chi d B + \sqrt3\chi^2 dA + \chi^3 d\cA\,.\label{pot2}
\eea
We may then derive all the equations of motion and Bianchi identities of
the original theory (\ref{d4lag}) from the doubled Lagrangian
\bea
{\cal L}_4 &=& R\, {*\oneone} - \ft32 {*d\phi}\wedge d\phi +
           \ft32 e^{2\phi} {*d\chi}\wedge d\chi  +
  \ft14 e^{-3\phi} {*\cF}\wedge \cF - \ft14 e^{3\phi} {*\cG}\wedge \cG\nn\\
&& -\ft14 e^{-\phi} {*F}\wedge F + \ft14 e^\phi {*G}\wedge G\,,\label{double}
\eea
together with (\ref{duals}).

   To make the $SL(2,\R)$ symmetry manifest, we define the $SL(2,\R)/O(1,1)$
scalar coset representative ${\cal V}$ and the matrix ${\cal M}$
\be
 {\cal V}=\begin{pmatrix} e^{\fft12\phi} & e^{\fft12\phi} \chi \\
                         0& e^{-\fft12\phi} \end{pmatrix}\,,\qquad
{\cal M}= {\cal V}^T\eta{\cal V}\,,\qquad
  \eta = \begin{pmatrix} -1&0 \\
                           0&1 \end{pmatrix}\,,
\ee
and we also define the $SL(2,\R)$ quartet of 1-form potentials 
${\bf A}^{\alpha\beta\gamma}$, which is
symmetric in its three $SL(2,\R)$ doublet indices $\alpha$,
$\beta$ and $\gamma$, by
\be
{\bf A}^{111}= \cA\,,\qquad {\bf A}^{112}= -\fft1{\sqrt3}\, A\,,\qquad
{\bf A}^{122}= \fft1{\sqrt3}\, B\,,\qquad {\bf A}^{222}= -\cB\,.
\ee

   It can then be seen that the Lagrangian (\ref{double}) may be written as
\be
{\cal L}= R\, {*\oneone} +\ft34 {\rm tr}({*d{\cal M}^{-1}}\wedge
 d{\cal M}) - \ft14 {*d{\bf A}^{\alpha\beta\gamma}}\wedge 
   d{\bf A}^{\delta\sigma\tau}\, ({\cal M}^{-1})_{\alpha\delta}\,
 ({\cal M}^{-1})_{\beta\sigma}\,({\cal M}^{-1})_{\gamma\tau}\,.
\label{lagsl2r}
\ee
Defining 
\be
{\bf F}^{\alpha\beta\gamma} \equiv ({\cal V}^{-1})_\delta{}^\alpha\,
  ({\cal V}^{-1})_\sigma{}^\beta\, ({\cal V}^{-1})_\tau{}^\gamma\,
   d {\bf A}^{\delta\sigma\tau}\,,
\ee
the duality relations (\ref{duals}) can be written as ${\bf F}^{222}=
  -{*{\bf F}^{111}}$ and ${\bf F}^{122} = -{*{\bf F}^{112}}$. They can
also be recast as
\be
d{\bf A}^{\alpha_1\alpha_2\alpha_3} = - {\cal M}^{\alpha_1\beta_1}\,
{\cal M}^{\alpha_2 \beta_2}\, {\cal M}^{\alpha_3\beta_3}\, 
\ep_{\beta_1\gamma_1}\, \ep_{\beta_2\gamma_2}\, \ep_{\beta_3\gamma_3}\,
      {*d{\bf A}^{\gamma_1\gamma_2\gamma_3}}\,.\label{dual2}
\ee

   The Lagrangian (\ref{lagsl2r}) and the duality relation
(\ref{dual2}) are manifestly invariant under the $SL(2,\R)$ transformations
\be
{\cal M} \longrightarrow \Lambda^T {\cal M} \Lambda\,,\qquad
{\bf A}^{\alpha\beta\gamma}\longrightarrow \Lambda_\delta{}^\alpha\,
 \Lambda_\sigma{}^\beta\, \Lambda_\tau{}^\gamma\, 
{\bf A}^{\delta\sigma\tau}\,,
\ee
where
\be
\Lambda=\begin{pmatrix} a & b \\
                         c & d\end{pmatrix}\,,\qquad ad-bc=1\,.\label{Lam}
\ee

   We are interested in the action of the $O(1,1)$ subgroup of
$SL(2,\R)$ transformations
\be
\Lambda = \begin{pmatrix} c & s \\
                          s & c\end{pmatrix}\,,
\ee
where $c\equiv \cosh\delta$ and $s\equiv \sinh\delta$, on the fields
$\phi$, $\chi$, $\cA$ and $A$ of the dimensionally-reduced original
solution.  The $O(1,1)$-transformed (primed) fields are therefore given by
\bea
\e^\phi &\longrightarrow& e^{\phi'} =e^\phi (c+s\chi)^2 - s^2 e^{-\phi}\,,\nn\\
\chi &\longrightarrow& \chi'=\fft{e^\phi (c+s\chi)(s+c \chi) - sc e^{-\phi} }{
                         e^\phi (c+s\chi)^2 - s^2 e^{-\phi} }\,,\nn\\
\cA &\longrightarrow& \cA'=c^3 \cA - \sqrt3 c^2 s A + \sqrt3 c s^2 B - s^3 \cB
                       \,,\nn\\
A &\longrightarrow& A'= -\sqrt3 c^2 s \cA + c(3c^2-2) A + (1-3c^2) s B
        + \sqrt3 c s^2 \cB\,.\label{transfields}
\eea
Having obtained the transformed four-dimensional fields, we then substitute
these quantities into (\ref{metans}) and (\ref{Aans}) in order to obtain 
the new solution of the minimal five-dimensional supergravity equations.
(We have not listed the transformed dual potentials $B$ and $\cal B$ in
(\ref{transfields}), since they are not needed in the lifting back to $D=5$.) 

   In section \ref{dipsec}, we construct a new dipole-charged solution of 
minimal
five-dimensional supergravity, and then in section \ref{elecsec}, we
apply the procedure we have just described in order to construct an
electrically charged generalisation.

\section{Solutions with Dipole Charge}\label{dipsec}

   In this section, we construct a new solution of the equations of motion 
of five-dimensional minimal supergravity, which generalises the 
charged Myers-Perry solution that is contained in the results of 
\cite{cvetyoum,cclp}.    The new solution that we construct here is 
analogous to
the generalisation of the neutral five-dimensional Myers-Perry black hole
that we obtained in \cite{lumeipope}.  

   Although the solution we shall construct in this section arises as
a generalisation of the electrically-charged version of the Myers-Perry
solution, it will be natural, from the point of view we are adopting here,
to reinterpret the ignorable coordinates in such a way that it
carries not an {\it electric} charge, but instead a magnetic {\it dipole}
charge.   This new dipole-charged solution will then itself be 
subjected, in section 4, to the ``charging'' procedure described in section 2.

    We take as our starting point the electrically-charged rotating
black hole solution of five-dimensional minimal supergravity, which
we find can be written as 
\bea
ds^2 &=& (x-y)\Big(\fft{dx^2}{4X} - \fft{dy^2}{4Y}\Big) -
      \fft{X (dt+y d\phi)^2}{x(x-y)} + \fft{Y(dt+x d\phi)^2}{y(x-y)}\nn\,,\\
&&- \fft{1}{xy} \left[\Big(\mu -\fft{q y}{x-y}\Big) dt 
 + (x+y)\Big(\mu - \fft{q y^2}{x^2-y^2}\Big) d\phi 
       +xy d\chi\right]^2\,,\nn\\
A &=& \fft{\sqrt3\, q}{(x-y)}\, (dt + y d\phi)\,,\label{cvsimp}
\eea
where
\be
X= (\mu+q)^2 + a_3\, x + a_2\, x^2\,,\qquad 
Y= \mu^2 + a_1\, y + a_2\, y^2\,.
\ee 
This is actually a special case of the charged rotating black holes found in
\cite{cvetyoum}, in which the three $U(1)$ charges are set equal. The form in
which the solution is written in (\ref{cvsimp}) is, however, considerably 
simpler.  If the charge parameter $q$ is set to zero, this solution reduces 
to the five-dimensional Myers-Perry \cite{myper} rotating black hole, 
in the simpler form introduced in \cite{chenlupope1}.  

   Our strategy now is analogous to the one we followed in \cite{lumeipope},
in which we obtained a generalisation of the $q=0$ five-dimensional rotating
black hole by introducing a suitable conformal factor multiplying a
four-dimensional subspace (the projection of the metric in (\ref{cvsimp})
that is orthogonal to $\del/\del t$), and at the same time adjusting the
detailed forms of the metric functions $X(x)$ and $Y(y)$.  In order
to do this, it will be convenient first to make the coordinate redefinitions
\be
t\longrightarrow \phi\,,\quad \phi\longrightarrow \psi\,,\quad
\chi\longrightarrow t\,,\quad x\longrightarrow \fft1{x}
\ee
in the solution (\ref{cvsimp}).  

   Introducing a conformal factor, and imposing the equations of motion
following from (\ref{d5lag}),
we obtain the new solution
\bea
ds^2\!\!\! &=&\!\!\! \fft{1}{(x-y)^2} \Bigr[ x(1-xy)\Big(
\fft{dx^2}{4G(x)} - \fft{dy^2}{4G(y)}\Big)  
  -\fft{G(x)(d\phi + y d\psi)^2}{1-xy} +
\fft{x G(y) (d\psi + x d\phi)^2}{y(1-xy)}\Bigr]\nn\\
&&-\fft{y}{x} \Big[dt + \fft{x}{y} \Big(\mu - \fft{q x y}{1-x y}\Big) d\phi +
             (x +y^{-1}) \Big(\mu -\fft{q x^2 y^2}{1-x^2y^2}\Big) d\psi\Big]^2
\,,\nn\\
A &=& \fft{\sqrt3 q x}{1-xy} (d\phi + y d\psi)\,,\label{gendipmet}
\eea
where
\be
G(\xi) = \mu^2 + a_1 \xi + a_2 \xi^2 + a_3 \xi^3 +
(\mu + q)^2 \xi^4\,.\label{genG}
\ee

  In appendix \ref{dipapp}, 
we shall discuss a limit of this new solution in which
it reduces to a black ring.  We also note here, in passing, that we can
take a different limit in which the solution reduces again to the 
charged Myers-Perry black hole.
In this limit, the $x$ and $y$ coordinates in (\ref{gendipmet}) are 
subjected to inverse scaling transformations, with
\bea
&&x\longrightarrow \fft1{\ep^2 x}\,,\quad y\longrightarrow \ep^2 y\,,\quad
\phi\longrightarrow \ep^{-1} t\,,\quad t\longrightarrow \ep^{-2} \chi\,,\quad
\psi\longrightarrow \phi\,,\nn\\
&&\mu\longrightarrow \ep^3 \mu\,,\quad q\longrightarrow \ep^3 q\,,\quad
a_1\longrightarrow \ep^4 a_1\,,\quad a_2\longrightarrow \ep^2 a_2\,,\quad
a_3\longrightarrow \ep^4 a_3\,,
\eea
and $\ep$ then sent to zero.  The solution reduces to (\ref{cvsimp}).

\section{Electrically Charged Dipole Solution}\label{elecsec}

      Having obtained the local solution (\ref{gendipmet}) 
that is supported by purely
magnetic dipole-like charges, we can apply the procedure
described in section 2 in order to introduce electric charge.
This is achieved by first reducing the solution on the
time direction, performing an $O(1,1)$ U-duality transformation, and then
lifting the solution back to $D=5$.  By this means we obtain the new solution
\bea
ds_5^2 &=& H ds_4^2 - \fft{y}{x H^2} (dt + \omega)^2\,,\nn\\
\omega &=& -\fft{q (s + cx)^3 (d\phi + y d\psi)}{x (1-x y)} +
\fft{(\mu+q)s^3 ((1+x y) d\phi + yd\psi)}{x}\nn\\
&& + \fft{c^3\mu ( (1+ xy) d\psi + x d\phi)}{y}\,,\nn\\
A&=&\fft{\sqrt3}{xH} \Big(cs\, (x-y) dt +
\fft{q(s + c x)^2(c + s y) (d\phi+y d\psi)}{(1-xy)}\nn\\
&& -
(\mu+q)cs^2((1+x y)d\phi +  y d\psi) -
\mu c^2 s ((1+x y) d\psi + x d\phi)\Big)\,,\nn\\
H&=& c^2 - \fft{s^2 y}{x}\,,\label{gensol}
\eea
where
$ds_4^2$ is the same as the four-dimensional base metric in (\ref{gendipmet}),
namely
\be
ds_4^2= \fft{1}{(x-y)^2} \Bigr[ x(1-xy)\Big(
\fft{dx^2}{4G(x)} - \fft{dy^2}{4G(y)}\Big)
  -\fft{G(x)(d\phi + y d\psi)^2}{1-xy} +
\fft{x G(y) (d\psi + x d\phi)^2}{y(1-xy)}\Bigr]\,,
\ee
and the function $G(\xi)$ is again given by (\ref{genG}):
\be
G(\xi) = \mu^2 + a_1 \xi + a_2 \xi^2 + a_3 \xi^3 +
(\mu + q)^2 \xi^4\,.\label{genG2}
\ee

\section{Global Analysis}\label{globalsec}

In this section, we shall discuss the global properties of
the new charged solutions (\ref{gensol}) that we have obtained.
It is advantageous to reparameterise the constants in the function $G(\xi)$ 
so that it is written as
\be
G(\xi)=(\mu + q)^2 (\xi-\xi_1) (\xi -\xi_2) (\xi-\xi_3)(\xi-\xi_4)\,,
\ee
with $\xi_1\xi_2\xi_3\xi_4=\mu^2/(\mu+q)^2$.   Furthermore, we
let $\xi_1<\xi_2<0<\xi_3<\xi_4$, with $\xi_1\xi_2<1$.  As we shall see
below, there exist black hole solutions in which the region outside the 
horizon is covered by the coordinate ranges 
\be
\xi_1\le x\le \xi_2\,,\qquad \xi_2\le y\le \xi_3\,.\label{endpoints}
\ee
Asymptotic infinity is located at $x=\xi_2=y$, and the horizon is
at $y=\xi_3$, with an ergosphere at $y=0$.  For
later convenience we shall parameterise the two negative roots $\xi_1$ and 
$\xi_2$ in terms of the positive constants
\be
\eta_1\equiv -\xi_1\,,\qquad \eta_2\equiv -\xi_2\,.\label{etaxi}
\ee

   The global analysis that we shall give in this section closely parallels
the one that we described in \cite{lumeipope} where new uncharged 
five-dimensional black holes were obtained.

   It is useful at this stage to shift the time coordinate
$t$ in (\ref{gensol}) according to
\be
t\rightarrow t - \Big(3c^2s q - c^3 (\mu+q) (\eta_1+\eta_2) +
\fft{\mu s^3}{\eta_1\eta_2}\Big) \psi +
\Big(3cs^2 q - c^3 (\mu+q) \eta_1\eta_2 + \fft{\mu s^3 (\eta_1 +\eta_2)}{
\eta_1\eta_2}\Big)\phi\,.\label{tshift}
\ee
In order to avoid naked CTCs, we find that the parameters in the
solution should be chosen so that
\be
q=\fft{\mu (s^3+ c^3 \eta_1) (1-\eta_1\eta_2)(1-\eta_2^2)}{
\eta_1\eta_2 (c^3 (\eta_1 + \eta_2 - \eta_1\eta_2^2) -
s (3c^2 - 3c s \eta_2 + s^2 \eta_2^2))}\,.
\ee
Note that if we turn off $q$, we can have either $\eta_1\eta_2=1$
or $\eta_1=-s^3/c^3$.  The former case leads to a charged rotating
black hole solution, whilst in the latter case, a conical singularity
cannot be avoided for finite $s$ or $c$, since $\eta_1$ would be less than
zero, as will become apparent presently.

   The analysis of the global properties of the solution rests upon 
investigating the behaviour at the singular points of the metric, which
occur when $x$ or $y$ approach their endpoints (see (\ref{endpoints})).  
We shall follow the method of analysis that was introduced in 
\cite{cvlupapo}, which begins by writing down the (appropriately
normalised) Killing vectors whose norms tend to zero at the spacelike
degeneration surfaces.  In the present case, these degenerations occur at
$x=\xi_1$, $x=\xi_2$ and $y=\xi_2$.  The associated Killing vectors
will be normalised so that they have unit Euclidean surface gravity.
(The Euclidean surface gravity $\kappa_E$ of a Killing vector $K$ whose
norm goes to zero on a spacelike degeneration surface is given by
$\kappa_E^2 = {\rm lim} (g^{\mu\nu} (\del_\mu K^2)(\del_\nu K^2)/(4K^2))$.)

   A redefinition of azimuthal coordinates is helpful at this stage.
We introduce $\phi_1$ and $\phi_2$, in place of $\phi$ and $\psi$,
defined by 
\bea
&&\psi= \nu (\eta_2 \phi_1 - \phi_2)\,,\qquad
\phi=\nu (\phi_1 - \eta_2\phi_2)\,,\nn\\
&& \nu\equiv \fft{\sqrt{\eta_2}}{(\mu+q)^2 (\eta_2 + \xi_3)(\eta_2 + \xi_4)
(\eta_1-\eta_2)}\,.
\eea
(Recall that $\eta_1$ and $\eta_2$ are defined by (\ref{etaxi}).)  We then
find that the normalised Killing vectors at the three degeneration 
surfaces are given by
\bea
x=\xi_1=-\eta_1:&&\ell_1=\alpha\Big( (1 - \eta_1\eta_2)\fft{\del}{\del\phi_1}
- (\eta_1-\eta_2)\fft{\del}{\del\phi_2}\Big)\,,\nn\\
y=\xi_2=-\eta_2:&&\ell_2=\fft{\del}{\del\phi_2}\,,\nn\\
x=\xi_2= -\eta_2:&&\ell_3=\fft{\del}{\del\phi_1}\,,\label{3kvs}
\eea
where the constant $\alpha$ is given by
\be
\alpha = \fft{\sqrt{\eta_1} (\eta_2 + \xi_3)(\eta_2+\xi_4)}{
\sqrt{\eta_2} (\eta_1 + \xi_3)(\eta_1 + \xi_4) (1-\eta_2^2)}\,.
\ee
Because of the chosen normalisation, if each Killing vector $\ell_i$ is 
written in terms of a coordinate $\psi_i$ as $\ell_i=\del/\del\psi_i$, then
advancing $\psi_i$ by an interval $2\pi$ will generate one complete
rotation around the origin in the plane of the degeneration (just like
the standard azimuthal angle in polar coordinates on the Euclidean 2-plane).

   The three Killing vectors (\ref{3kvs}) span a two-dimensional vector
space, and so there is a linear relation between them.  A necessary condition
for avoiding conical singularities is that the coefficients
in this linear relation must be rationally related, since otherwise it
would be possible, by taking integer combinations of $2\pi$ rotations 
around the circles, to generate a translation that implied an identification
of arbitrarily close points on the manifold.  By an overall scaling in the
linear relation, we may therefore say, without loss of generality, that
the coefficients must be coprime integers $m$, $n$ and $p$, 
\be
p\, \ell_3 = m \, \ell_1 + n\, \ell_2\,.
\ee

   The Killing vectors $\ell_1$ and $\ell_2$ can degenerate simultaneously,
and, as discussed in \cite{cvlupapo}, this
implies an additional condition, leading to the constraint
\be
\ell_3=m\,\ell_1 + n\,\ell_2\,.
\ee
It therefore follows from (\ref{3kvs}) that the coprime integers 
$(m,n)$ are given by
\be
m=\fft{1}{\alpha (1-\eta_1\eta_2)}\,,\qquad
n=\fft{\eta_1-\eta_2}{1-\eta_1\eta_2}\,.
\ee
Note that we therefore have
\be
\ell_1= \fft1{m}\, \fft{\del}{\del\phi_1} - \fft{n}{m}\, 
         \fft{\del}{\del\phi_2}\,.
\ee

   It is helpful to look at the double degeneration, at $x=\xi_1$ and 
$y=\xi_2$, in more detail.  Introducing new coordinates $\rho$ and 
$\vartheta$, we write 
\be
x= -\eta_1 + e_1\, \rho^2 \sin^2\vartheta\,,\qquad
y= -\eta_2 + e_2 \, \rho^2 \cos^2\vartheta\,,
\ee
where
\be
e_1 = (\eta_1+\xi_3)[\mu^2+ (\mu+q)^2\, \eta_1^2\, \eta_2\,\xi_3]\,,\quad
e_2= (\eta_2+\xi_3)[\mu^2+ (\mu+q)^2\, \eta_1\, \eta_2^2\,\xi_3]\,,
\ee
and then look at the metric (\ref{gensol}) in the limit when $\rho$ is 
small.  It is convenient first to make a further change of azimuthal
coordinates, introducing $\chi_1$ and $\chi_2$  given by
\be
\chi_1= m \phi_1\,,\qquad \chi_2 = \phi_2 + n \phi_1\,.\label{chiphi}
\ee
We then find that near $\rho=0$, the metric (\ref{gensol}) approaches
\bea
ds^2 &=&
\fft{\eta_1\, \eta_2\, (1-\eta_1\, \eta_2)(c^2 \eta_1-s^2\eta_2)\xi_3}{
(\eta_1-\eta_2)^3}\, \Big( d\rho^2 +\rho^2[d\vartheta^2 + \sin^2\vartheta
   d\chi_1^2 + \cos^2\vartheta d\chi_2^2]\Big)\nn\\
&&-\fft{\eta_1\, \eta_2}{(c^2 \eta_1 - s^2 \eta_2)^2} dt^2\,.
\eea
This shows that to avoid the occurrence of conical singularities,
$\chi_1$ and $\chi_2$ must have independent $2\pi$ periods (so that
the collapsing $\rho=$constant surfaces are 3-spheres with no 
identifications).   Thus $\chi_1$ and $\chi_2$ are periodic on a
square lattice of side $2\pi$. 

   One might think that there would be another double degeneration
surface at $x=y=\xi_2$, leading to another restriction on the periods,
but in fact this region actually describes asymptotic infinity.   
 To make thid manifest, we introduce new coordinates $r$ and $\theta$,
in terms of which we write
\be
x=-\eta_2 - \fft{\sqrt{\eta_2}(1-\eta_2^2)\nu\,\cos^2\theta}{r^2}\,,\qquad
y=-\eta_2 + \fft{\sqrt{\eta_2}(1-\eta_2^2)\nu\,\sin^2\theta}{r^2}\,.
\ee
In the limit $r\rightarrow \infty$, we then find that 
the metric (\ref{gensol}) approaches
\be
ds^2=-dt^2 + dr^2 + r^2 (d\theta^2 + \cos^2\theta\, d\phi_1^2 +
\sin^2\theta\, d\phi_2^2)\,.\label{metinf}
\ee

    Since from (\ref{chiphi}) we have
\be
\phi_1= \fft1{m}\, \chi_1\,,\qquad \phi_2= \chi_2 - \fft{n}{m}\, \chi_1\,,
\ee
and since we have already determined that $\chi_1$ and $\chi_2$ are periodic
on a square lattice of side $2\pi$, it follows that $\phi_1$ and $\phi_2$ 
are periodic on as tilted lattice, which may be defined by the
two identifications
\bea
1)&&\qquad \phi_1\longrightarrow \phi_1\,,\qquad \phi_2\longrightarrow
   \phi_2 + 2\pi\,,\nn\\
2)&&\qquad \phi_1\longrightarrow \phi_1 + \fft{2\pi}{m}\,,\qquad 
  \phi_2\longrightarrow \phi_2 - \fft{2\pi n}{m}\,.\label{phiident}
\eea
These identifications imply that the $r=\hbox{constant}$ 
surfaces in (\ref{metinf})
are the lens space $L(m;n)$.  This is defined by considering complex
coordinates $(z_1,z_2)$, in terms of which the unit $S^3$ is described by
$|z_1|^2 + |z_2|^2=1$.  The sphere is then quotiented by the 
identifications
\be
(z_1,z_2) \longrightarrow (z_1\, e^{2\pi \im/m}, z_2\, e^{2\pi\im n/m})\,,
\label{lensdef}
\ee
for coprime integers $m$ and $n$ with $1\le n\le m-1$, to define the
lens space $L(m;n)$.  We may take (\ref{lensdef}) to define the discrete
subgroup $\Gamma(m;n)$ of $SO(4)$.
By taking $z_1=\sin\theta e^{\im\phi_1}$ and
$z_2=\cos\theta e^{-\im\phi_2}$, it can be seen that (\ref{lensdef})
implies (\ref{phiident}).  In turn, this implies that the $r=\hbox{constant}$
surfaces at large $r$ are lens space $L(m;n)= S^3/\Gamma(m;n)$.

   The event horizon is located at $y=\xi_3$.  In order to study the geometry
and topology of the horizon, it is helpful to introduce a yet further new
pair of azimuthal coordinates $\td\phi_1$ and $\td\phi_2$, defined by
\be 
\td\phi_1= \fft1{n}\, \chi_2 = \phi_1 + \fft{1}{n}\, \phi_2\,,\qquad
  \td\phi_2 = \chi_1 - \fft{m}{n}\, \chi_2 = -\fft{m}{n}\, \phi_2\,.
\ee
In terms of these, the Killing vectors $\ell_1$ and $\ell_3$ in (\ref{3kvs}),
which degenerate at $x=\xi_1$ and $x=\xi_2$ respectively, are simply given 
by
\be 
\ell_1 = \fft{\del}{\del\td\phi_2}\,,\qquad 
 \ell_2= \fft{\del}{\del\td\phi_1}\,.
\ee
Writing $x=\xi_1+ \rho_1^2$ or $x=\xi_2-\rho_2^2$, for small $\rho_i$, 
to describe the
regions near the two degeneration surfaces, we find that the metric on 
the horizon has the form
\be
ds_H^2 \sim b_1 (d\rho_1^2 + \rho_1^2\, d\td\phi_2^2) + c_1 (d\td\phi_1+
   \cA_1)^2
\ee
near $x=\xi_1$ and
\be
ds_H^2 \sim b_2 (d\rho_2^2 + \rho_2^2\, d\td\phi_1^2) + c_2 (d\td\phi_2+
   \cA_2)^2
\ee
near $x=\xi_2$, where the $b_i$ and $c_i$ are non-singular and non-zero.
The 1-forms $\cA_i$ are of the form $\cA_1= \rho_1^2 f_1 d\td\phi_2 +
    g_1 dt$ and $\cA_2= \rho_2^2 f_2 d\td\phi_1 +
    g_2 dt$, where $f_i$ and $g_i$ are non-singular.  It is straightforward 
to see that, in a manner quite analogous to the situation for static
lens-space black holes in \cite{lumeipope}, the horizon therefore has
the local geometry of a distorted 3-sphere.  As in \cite{lumeipope}, 
the relation between $(\td\phi_1,\td\phi_2)$ and $(\chi_1,\chi_2)$ is
just like the relation between $(\phi_1,\phi_2)$ and $(\chi_1,\chi_2)$, except
that the roles of the integers $m$ and $n$ are reversed.  This means that
the identifications of the $\td\phi_i$ coordinates imply that the topology
of the horizon is $S^3/\Gamma(n;m)$, or, in other words, the lens
space $L(n;m)$.

  By studying the behaviour near infinity,
we can determine the mass and angular momenta by means of the
Komar integrals  $M=3/(32\pi) \int{*dK[t]}$ and
  $J_{\phi_i}=1/(8\pi) \int{*dK[\phi_i]}$, where $K[t]$ and $K[\phi_i]$ 
are the 1-forms associated with $\del/\del t$ and $\del/\del \phi_i$.
They, and the electric charge $Q_e=1/(16\pi)\int ({*F} -1/\sqrt3\, A\wedge F)$
are given by
\bea
M &=& \fft{3\pi (1 + 2s^2) (1-\eta_2^2)\nu}{8m\sqrt{\eta_2}}
\,,\qquad Q_e= \fft{\sqrt3\,\pi sc (1-\eta_2^2)\nu}{4m\sqrt{\eta_2}}\nn\\
J_{\phi_1} &=& \fft{\pi\nu^2}{4m\eta_2^{3/2}}\Big(
\mu (1-\eta_2^2)^2 (s^3 + c^3 \eta_2) +
q\,\eta_2^2 (3c^2 s - c (2 + 5 s^2) \eta_2 + s^3 \eta_2^2 +
c^3 \eta_2^3)\Big)\,,\nn\\
J_{\phi_2} &=&\fft{\mu\pi(1-\eta_2^2)\nu^2}{4m\eta_1\eta_2^{3/2}
(3c^2 s - 3 c s^2 \eta_2 + s^3 \eta_2^2 -c^3 (
\eta_1+\eta_2-\eta_1\eta_2^2))}\Big(c^3 s^3 (\eta_1-\eta_2)^2 \eta_2
\nn\\
&& + s^2\eta_2 (3c^4 \eta_1 - s^4\eta_2)(2-\eta_1\eta_2 -\eta_2^2)
+c^2 (c^4 \eta_1 - 3s^4\eta_2) (\eta_1 + \eta_2 -2\eta_1\eta_2^2)
\nn\\
&&-3cs (c^4 \eta_1 - s^4\eta_2) (1-\eta_1\eta_2^3)\Big)\,.
\eea
The horizon is located at $y=\xi_3$, which is topologically
a lens space $L(n;m)$.  The temperature, entropy and angular velocities
and electric potentials can be easily calculated, and are
given by
\bea
S &=& \fft{\pi^2(\eta_1-\eta_2)(1-\eta_2^2)\nu^2\beta}{
2m\sqrt{\xi_3} \eta_1\eta_2 (\eta_1+\xi_3)(\eta_2 +\xi_3)}\,,\nn\\
T &=& \fft{1}{2\pi\beta}
(\mu+q)^2 \sqrt{\xi_3} \, \eta_1\eta_2 (\eta_1 + \xi_3) (\eta_2 +\xi_3)
(\xi_4-\xi_3)\,,\nn\\
\Omega_{\phi_1} &=& \fft{\eta_1\eta_2\xi_3 (\xi_3+\eta_2)}{\nu\beta
(1-\eta_2^2)}\,,\qquad 
\Omega_{\phi_2} = \fft{\eta_1\eta_2\xi_3(1+\eta_2\xi_3)}{\nu\beta
(1-\eta_2^2)}\\
\Phi_e&=& \fft{\sqrt3\, \eta_1\, (\eta_2+\xi_3)}{\beta (1-\eta_2^2)}\Big(
\mu c s (c-s\xi_3)(1-\eta_2^2)(1 + \eta_2\xi_3)\nn\\
&&\qquad \qquad - q \eta_2\xi_3 [-c\eta_2 + s(2 + \eta_2^2) +
s^3 (3 + \eta_2^2) + c s^2 (\xi_3 -4\eta_2 - \eta_2^2\xi_3)]\Big)\,,
\nn
\eea
where
\bea
\beta &=& \mu (1 + \eta_1\xi_3)(1+\eta_2\xi_3)
(c^3\eta_1\eta_2 - s^3 \xi_3)\nn\\
&& -q \eta_1\eta_2\xi_3 \Big(3c^2s + 3 c s^2 \xi_3 + s^3 \xi_3^2 -
c^3(\eta_1+\eta_2 + \eta_1\eta_2\xi_3) \Big)\,,
\eea

\section{Conclusions}

   In this paper, we have constructed new charged solutions of 
five-dimensional supergarvity, which generalise the new Ricci-flat metrics
that we obtained in \cite{lumeipope}.  To do this, we first established a
procedure, utilising the global $SL(2,\R)$ symmetry of the four-dimensional
theory obtained from a timelike
reduction of five-dimensional minimal supergravity, which enabled us to
introduce electric charge in any stationary solution of the five-dimensional
theory.  This simple prodedure, described in section \ref{chargingsec}, is of
quite general utility, and has applications beyond those in this paper.

   We then constructed the ``seed'' solution in five-dimensional minimal
supergravity, which forms the starting point for the charging proceedure.
The starting point for the construction of the seed solution was 
the electrically-charged generalisation
of the five-dimensional rotatating Myers-Perry black hole; this is a 
specialisation (to minimal supergravity) of more general 
charged solutions obtained in \cite{cvetyoum}.  After a reinterpretation of
the coordinates, this solution can be viewed as carrying a magnetic dipole
charge, rather than an electric charge.  Next, we looked for a generalisation
of this solution, along the same lines as the generalisation of Ricci-flat
metrics that allowed us to construct new vacuum solutions in \cite{lumeipope}.
This involved the introduction of a conformal factor for a four-dimensional
base.  This provided us with new rotating solutions of minimal
five-dimensional supergravity, carrying magnetic dipole charge.  These
formed the seed solutions to which we then applied the charging procedure.

   After charging the seed solution, we arrived at our new solution
(\ref{gensol}), which carries both electric charge and magnetic dipole
charge.  A special limit of the solution gives back the charged
rotating black hole in five-dimensional minimal supergravity.  In
another limit, the solution reduces to a class of black rings found in
\cite{emparan,elvempfig}.  We discussed this limit in detail in
appendix \ref{dipapp}.

   In the remainder of the body of the paper, we analysed the global
proprties of the general solution (\ref{gensol}).  We found that provided
the parameters are chosen properly, with two algebraic conditions 
characterised by coprime integers $m$ and $n$, 
the solution describes a stationary
black hole spacetime that is asymptotically locally flat, with an horizon
that is topologically the lens space $L(n;m)=S^3/\Gamma(n;m)$.  At
large distance, the spacetime approaches (Minkowski)$_5/\Gamma(m;n)$.  
The two algebraic conditions involving $m$ and $n$ arose 
from requiring that the spacetime be free from conical singularities.

   Our results in this paper and in \cite{lumeipope}
have all been restricted to five dimensions, and to 
the case where there is no cosmological constant. Generalisations beyond 
five dimensions, and also generalisations with a cosmological constant or
a gauge coupling in supergravity, would be of considerable interest.  
Such solutions might admit limits that could describe
higher-dimensional black holes with ring-like topologies, or black
rings in asymptotically AdS backgrounds.

\section*{Acknowledgements}

      We are grateful to Gary Gibbons and Harvey Reall for helpful
discussions.  H.L. is grateful to ICTS at the University of Science
and Technology of China, and C.N.P. is grateful to the
Relativity and Cosmology groups at the Centre for Mathematical Sciences,
Cambridge, for hospitality during the course of this work.  
Research is supported in part by DOE grant DE-FG03-95ER40917.

\newpage
\centerline{\bf \Large Appendices}
\bigskip

\appendix

\section{3-Charge Solutions in the STU Model}

   The charged solution (\ref{gensol}) was obtained by applying the
charging procedure given in section \ref{chargingsec} to the 
solution (\ref{gendipmet}) that we obtained in section \ref{dipsec}. 
This procedure added electric charge to any stationary solution of
five-dimensional minimal supergravity, by means of an $O(1,1)$ duality
transformation of its timelike dimensional reduction to four dimensions.

   Another way of adding charge in five dimensions, which moreover allows
the introduction of three independent charges in the ${\cal N}=2$ 
STU supergravity theory with $U(1)^3$ gauge fields, involves a 
thrice-repeated sequence of
lifting to six dimensions, performing a Lorentz boost, and reducing again
to $D=5$.  In the intermediate stages, the discrete symmetries of the
STU model are used twice over, to transfer the first and second charges
from the Kaluza-Klein vector of the $D=6$ to $D=5$ reduction to one
of the other $U(1)$ fields.  We have implemented this procedure in 
order to construct the generalisation of the solution (\ref{gensol})
to the STU model.

   We find that the generalised solution, with three independent electric
charges, is given by
\bea
ds^2&=&(H_1H_2H_3)^{1/3}ds_4^2-\frac{xy(dt+\omega)^2}{(H_1H_2H_3)^{2/3}}\,,
\nonumber\\
ds_4^2&=&\frac{x-y}{(1-xy)^2}\left[\frac{dx^2}{4X}-\frac{dy^2}{4Y}
-\frac{X(d\phi+yd\psi)^2}{x(x-y)^2}+\frac{Y(d\phi+xd\psi)^2}{y(x-y)^2}\right
]\,,\nonumber\\
\omega&=&\frac{\mu c_1c_2c_3(d\phi+(x+y)d\psi)}{xy}-(\mu+q)s_1s_2s_3
(d\phi(x+y)+xyd\psi)\nonumber\\
&&-\frac{q(c_1-s_1x)(c_2-s_2x)(c_3-s_3x)(d\phi+yd\psi)}{x(x-y)}~;\nonumber\\
X&=&(\mu+q)^2+a_3x+a_2x^2+a_1x^3+\mu^2x^4~;\nonumber\\
Y&=&\mu^2+a_1y+a_2y^2+a_3y^3+(\mu+q)^2y^4~.
\eea
The gauge and scalar fields are given by
\bea
A_i&=&\frac{c_is_i(1-x y)dt+(\mu+q)c_is_js_k(d\phi(x+y)+xyd\psi)
-\mu s_ic_jc_k(d\phi+(x+y)d\psi)}{H_i}\nonumber\\
&&-\frac{q(c_i-s_iy)(c_j-s_jx)(c_k-s_kx)(d\phi+yd\psi)}{(x-y)H_i}\,,\nonumber
\\
H_i&=&c_i^2-s_i^2xy~;~~~i\neq j\neq k~~~{\rm and}~~~i,j,k=1,2,3\,.
\eea
Here, we have defined $c_i=\cos\delta_i$ and $s_i=\sinh\delta_i$, where 
$\delta_i$ are the three boost parameters for the three electric charges. 

   We have presented the solution in a symmetrical form for
the $x$ and $y$ coordinates.   The solution reduces to the 
solution (\ref{gensol}) in minimal five-dimensional supergravity if
the three charges are set equal (\ie $\delta_1=\delta_2=\delta_3=\delta$),
and also $x$ is sent to $1/x$.

\section{Black Ring Limit}\label{dipapp}

    In this appendix, we consider 
a limiting form of the general charged solutions
(\ref{gensol}), which gives rise to black ring spacetimes.\footnote{These
black ring solutions are contained within those found in \cite{elvempfig}.
We give a discussion for their global properties here because our methods
of analysis are rather different from those employed in \cite{elvempfig},
and also we give some explicit results for the thermodynamic properties
that did not appear in \cite{elvempfig}.}  This is 
obtained by making the rescalings 
\bea
&&x\rightarrow \epsilon^2\,x\,,\qquad y\rightarrow \epsilon^2\,y
\qquad \psi\rightarrow \epsilon\, \psi\,,\qquad
\phi\rightarrow \epsilon\, \phi\,,\nn\\
&& q\rightarrow \epsilon^{-3}\, q\,,\qquad
\mu\rightarrow \epsilon\,\mu\,,\qquad
a_i\rightarrow a_i\, \epsilon^{2i-2}\,, \qquad
 t\rightarrow t + 3 c s^2 q\ep^{-1}\, \phi
\,,\label{ringlimit2}
\eea
and then sending $\ep$ to zero.  In the process, we also
discard a divergent pure gauge term in the $U(1)$ potential.  This leads
to the solution
\bea
ds^2 &=& \fft{H}{(x-y)^2} \Big(\fft{x dx^2}{4 G(x)} -
\fft{x dy^2}{4 G(y)} - G(x) d\phi^2 +
\fft{x G(y) d\psi^2}{y}\Big) - \fft{y}{xH^2} (dt + \omega)^2\,,\nn\\
A &=& \fft{\sqrt3}{x H} \Big( cs (x-y) dt + (q s^3 y^2 -
c^2 s (\mu - 2 q x y)) d\psi + (c^3 q x^2 - c s^2 (\mu - 2 q x y))
d\phi\Big)\,,\nn\\
\omega&=&\Big(\fft{c^3\mu}{y} - 3 c s^2 q y\Big)d\psi +
       \Big(\fft{s^3\mu}{x} - 3 c^2 s q x\Big) d\phi\,,\label{cring}
\eea
where the function $H$ is the same as that given by (\ref{gensol}).
We again reparameterise so that $G(\xi)$ is given by
\be
G(\xi) = q^2 (\xi-\xi_1)(\xi-\xi_2) (\xi-\xi_3)(\xi-\xi_4)\,,
\ee
where the the four roots satisfy $\xi_1\xi_2\xi_3\xi_4=\mu^2/q^2$.
Furthermore we require that $\xi_1< \xi_2<0<\xi_3<\xi_4$.
Since the solution is obtained from a scaling limit, it follows that
there is a residual scaling symmetry which enables us to set,
without loss of generality, $\xi_2=-1$.  We introduce 
\be
\eta \equiv \sqrt{-\xi_1}>1\,.
\ee

     The $x$ and $y$ coordinates will be taken to lie in the ranges 
\be
-\eta^2 \le x \le -1\,,\qquad -1\le y \le \infty\,.
\ee
There is a region corresponding to asymptotic infinity located  
at $x=-1=y$.   Outer and inner
horizons are located at $y=\xi_3$ and $y=\xi_4$ respectively.
In order for $g_{\psi\psi}$ and $g_{\phi\phi}$ to be non-negative
on the degenerate surfaces $x=-\eta^2$, $x=-1$ and $y=-1$,
it is necessary to shift the time coordinate $t$, according to
\be
t\rightarrow t + c (c^2 \mu - 3 q s^2)\, \psi
                - s (3c^2 q - \mu s^2)\, \phi\,,
\ee
and also to impose the additional constraint
\be
s(\mu s^2 + 3 c^2 q \eta^2)=0\,.
\ee
This leads to a bifurcation of solutions:
\be
s=0\,,\qquad \hbox{or}\qquad q=-\fft{\mu\,s^2}{3c^2 \eta^2}\,.
\label{ringtwocase}
\ee
The first case corresponds to turning off the electric charge. 
The second case has a non-vanishing 
electric charge, related to the dipole charge.  We
shall discuss the two cases separately.

\subsection{Dipole ring}

This solution can be found in \cite{emparan} in a different coordinate
system.  In this case, we turn off the electric charges by setting
$s=0$, and hence $c=1$.  The Killing vector $\del/\del\phi$
degenerates at both $x=\xi_1\equiv -\eta^2$ and $x=\xi_2=-1$.  For the
two resulting periodicity conditions to be consistent, it is necessary
that the parameters in the solution be chosen such that
\be
q^2=\fft{\mu^2(\xi_3-\eta)}{\eta^3 \xi_3 (\xi_3 + 1 + \eta + \eta^2)}
\,.
\ee
Thus we see that $\xi_3\ge \eta$.  Furthermore, since we have $\xi_4\ge \xi_3$,
it follows that the parameter $\xi_3$ lies in the range
\be
\eta\le \xi_3\le \eta + (1+\eta)\sqrt{\eta}\,.\label{xi3range}
\ee
The lower bound corresponds to the neutral black ring solution, whilst
the upper bound corresponds to an extremal black ring.

        The Killing vector $\del/\del\psi$ degenerates at $y=-1$.  
This determines the periodicity of $\psi$.  Ii is convenient to
rescale both the azimuthal coordinates $\phi$ and $\psi$, according to
\be
\phi=\nu \td\phi\,,\qquad \psi=\nu \td\psi\,,
\ee
where
\be
\nu=\fft{\eta^3\xi_3 (1+\eta + \eta^2 + \xi_3)}{\mu^2
(\eta-1)(\eta+1)^2(\eta^2+\xi_3)(1+\xi_3)}\,.
\ee
Then, $\td\phi$ and $\td\psi$ have independent $2\pi$ periods (\ie they are 
defined on a square lattice of side $2\pi$).
With this condition, our solution is regular without naked CTCs outside
the horizon.

      The asymptotic region at infinity is located at $x=y=-1$.  This may
be seen by making the coordinate transformations
\be
x=-1 - \fft{\nu\,\cos^2\theta}{r^2}\,,\qquad
y=-1 + \fft{\nu\,\sin^2\theta}{r^2}\,.\label{asymtrans}
\ee
At large $r$, the metric then aproaches
\be
ds^2\sim -dt^2 + dr^2 + r^2 (d\theta^2 + \cos^2\theta d\td\phi^2 + \sin^2
\theta d\td\psi^2)\,.\label{asym1}
\ee
By evaluating Komar integrals at infinity, we find that the mass and 
the angular momenta are given by
\be
M=\ft38\pi\,\nu\,,\qquad J_{\td\psi}=\ft14\pi\mu\,\nu^2\,,\qquad
J_{\td\phi}=0\,.
\ee

      The horizon is located at $y=\xi_3$.  The Killing vector
\be
\ell=\fft{\del}{\del t} +\Omega_{\td\psi}\, \fft{\del}{\del\td \psi}
\ee
becomes null on the horizon,
where $\Omega_{\td\psi}$, the angular velocity at the horizon, is 
given by
\be
\Omega_{\td\psi} = -\fft{\xi_3}{\mu\nu (1+\xi_3)}\,.
\ee
By calculating the surface gravity $\kappa$ at the horizon, we find that
the Hawking temperature $T=\kappa/(2\pi)$ is given by 
\be
T=\fft{\mu(\xi_3+ \eta^2)\Big(\eta(1+\eta)^2 -(\xi_3-\eta)^2\Big) 
                     }{2\pi \eta^3\sqrt{\xi_3}
(1+\eta+\eta^2 +\xi_3)}\,,
\ee
It is straightforward to see that the horion has the topology $S^2\times S^1$,
and that the entropy, equal to one quarter of the horizon area, is
\be
S=\fft{\pi^2 \mu \nu^2 (\eta^2-1)}{2\sqrt{\xi_3} (\eta^2 + \xi_3)}\,.
\ee

     The solution is supported purely by a magnetic dipole charge, given by
\be
{\cal D} =\ft18 \int F = \ft14\pi\sqrt3\, q\nu\ (\eta^2-1)\,,
\ee
where the integration is performed over the $S^2$ component of the horizon.
The potential difference between the horizon and infinity is calculated by
Hodge dualising the magnetic 2-form $F$ to an electric 3-form $G={*F}=
 dB + A\wedge F/\sqrt3$.  The dipole potential $\Phi_{\cal D}$ is given by
the difference between $B_{\td\psi t}$ at the horizon and infinity, and so
\be
\Phi_{\cal D}=\sqrt3 q\,\nu (\xi_3 + 1)\,.
\ee
With these quantities, we find that the first law of thermodynamics is
satisfied.  In particular, we have
\be
dM=TdS + \Omega_\psi dJ_\psi + \Phi_{\cal D} \, d{\cal D}\,,\qquad
M=\ft32 T S + \ft32 \Omega_\psi J_\psi + \ft12 \Phi_{\cal D}
{\cal D}\,.
\ee
(See \cite{cophor} for a detailed discussion of the role of 
magnetic dipole charge in black hole thermodynamics.)

        There exists an extremal limit when $\xi_3=\xi_4$, which
implies that
\be
\xi_3=\eta + (1+\eta)\sqrt{\eta}\,.
\ee
In this limit, the temperature goes to zero, and the near horizon
geometry becomes a warped product of a distorted AdS$_3$ and
$S^2$.  There is a decoupling limit which reduces the solution
to its near horizon geometry, analogous to the AdS$_5\times S^5$ 
decoupling limit of the D3-brane.  Letting $y=\xi_3 - \epsilon\, r$
and $t\rightarrow t/\epsilon$, and sending $\epsilon$ to zero, we
have
\bea
ds^2&=&\fft{\eta^2 \xi_3\, x\, dx^2}{4\mu^2 (x+1)(x+\eta^2)(x-\xi_3)^4}
-\fft{\eta^2\xi_3^2(x+1)(x+\eta^2) d\phi^2}{\mu^2(\sqrt{\eta}-1)^2
(1+\xi_3)^6}\nn\\
&&
-\fft{\eta^3\xi_3^2}{4\mu^2(1+\xi_3)^4x}\Big(
d\psi'  + r dt'\Big)^2
-\fft{\eta^2\xi_3^2\, x}{4\mu^2(1+\xi_3) (\eta^2+\xi_3)(x-\xi_3)^2}
\Big(\fft{dr^2}{r^2} + r^2 dt'^2\Big)\,,\nn\\
A&=&\fft{\sqrt3\,\eta\xi_3\, x}{\mu(\sqrt{\eta}-1)(1+\xi_3)^3}\,
d\phi\,,\label{ads3}
\eea
where
\be
t'=\fft{2\mu(\eta^2+\xi_3)}{\eta^2 \xi_3^{3/2}}\, t\,.\qquad
\psi'=\fft{\sqrt{\eta}-1}{2\sqrt{\eta\,\xi_3}}(\psi - \Omega_{\psi} t)\,.
\ee
The first two terms in the metric in (\ref{ads3}) give rise
to an $S^2$, whilst the remaining terms describe a homogeneously
squashed AdS$_3$, viewed as a $U(1)$ bundle over AdS$_2$.

\subsection{Electrically-charged dipole ring}

This solution can be found in \cite{elvempfig} in a different
coordinate system.  The solution corresponds to the second choice in
(\ref{ringtwocase}), which requires that the parameters determining
the electric charge and the dipole charge are related by
\be
q=-\fft{\mu s^2}{3c^2 \eta^2}\,.\label{qcon}
\ee
The compatibility between the periodicity conditions on $\phi$ at 
the two degeneration surfaces $x=-\eta^2$ and $x=-1$
implies that the parameters must be chosen such that
\be
\fft{(\eta^2 +\xi_3) (\eta^2+\xi_4)}{\eta (1 +\xi_3)
(1+\xi_4)}=1\,.\label{ringcon}
\ee
It is convenient again to rescale the azimuthal coordinates $\phi$ and $\psi$, 
according to
$\phi=\nu\, \td\phi$ and $\psi=\nu\, \td\psi$, where now
\be
\nu=\fft{1}{q^2(\eta^2-1)(1+\xi_3)(1+\xi_4)}\,.\label{cringnu}
\ee
We then find that $\td\phi$ and $\td \psi$ have independent $2\pi$ periods,
and are thus defined on a square lattice of side $2\pi$.

    Asymptotic infinity is at $x=y=-1$.  To see this
explicitly, we can make the same coordinate transformation
(\ref{asymtrans}), with $\nu$ now given by (\ref{cringnu}). At large $r$
the metric again approaches (\ref{asym1}).
From the aymptotic form of the solution we find that 
the mass $M$, electric charge $Q_e$ and angular momenta $J_{\td\phi}$ and
$J_{\td\psi}$ are given by
\bea
M&=&  \ft38 \pi (1 + 2s^2)\nu \,,\qquad Q_e= \ft14 \sqrt3 \pi s c \nu\,,\nn\\
J_{\td\psi}&=&\ft14 \pi c(\mu c^2 + 3 q s^2)\nu^2\,,\qquad
J_{\td\phi}=\ft14 \pi s(\mu s^2 + 3 q c^2)\nu^2\,.
\eea

     The horizon is located at $y=\xi_3$, and the corresponding
Killing vector that vanishes on the horizon is given by
\be
\ell=\fft{\del}{\del t} - \Omega_{\td\psi} \fft{\del}{\del\td\psi}\,,
\ee
where $\Omega_{\td \psi}$ is the angular velocity, given by
\be
\Omega_{\td \psi} =  \fft{\xi_3}{\nu c (1+\xi_3) (
\mu c^2 - 3 q s^2 \xi_3)}\,.
\ee
(Note that $\Omega_{\td\phi}=0$, even though $J_{\td\phi}$ is non-zero.)
The temperature is given by
\be
T=\fft{q^2 \sqrt{\xi_3} (\eta^2 + \xi_3) (\xi_4-\xi_3)}{
2\pi\, c (c^2 \mu - 3q s^2 \xi_3)}\,.
\ee
The horizon can be seen to have the topology $S^2\times S^1$, and by
calculating its area, we find that the entropy is
\be
S= \fft{c \nu^2  \pi^2 (\eta^2-1)(c^2\mu - 3qs^2 \xi_3)}{2\sqrt{\xi_3}
 (\eta^2+\xi_3)}\,.
\ee
The electrostatic potential difference between horizon and infinity is
given by
\be
\Phi_e=\fft{\sqrt3\, s (q \xi_3 + c^2 (\mu-3q \xi_3))}{
c (3q \xi_3 + c^2 (\mu-3q \xi_3))}\,.
\ee

   The magnetic dipole charge is given by
\be
{\cal D} = \ft18 \int F = \fft{3\sqrt3\, \pi s^2 c^3 \eta^2\xi_3}{
   2\mu (1+\xi_3)(9c^4 \eta^2 + s^4 \xi_3)}\,.
\ee
This is related to $J_{\td\phi}$ and the electric charge $Q_e$ by
\be
J_{\td\phi} = \fft{2}{\pi}\, Q_e\, {\cal D}\,.
\ee
From the expected generalised Smarr relation 
\be
M=\ft32 T S + \ft32 \Omega_{\td \psi} J_{\td \psi} + \Phi_e Q_e +
\ft12 \Phi_{\cal D} {\cal D}\,,\label{edipole}
\ee
we find that the dipole potential is given by
\be
\Phi_{\cal D} = \fft{\sqrt3\, s^2 c\,  \eta^2 \xi_3}{2\mu(\eta^2-1)}\Big[
 \fft3{9c^4 \eta^2 + s^4 \xi_3} -\fft1{c^4\eta^2 + s^4 \xi_3}\Big]\,.
\ee
It can then be verified that the first law
\be
dM=T dS + \Omega_{\td\psi} dJ_{\td\psi} + \Phi_e dQ_e +
\Phi_{\cal D} d{\cal D}
\ee
is satisfied.


\begin{thebibliography}{99}

\bm{empreal} R. Emparan and H.S. Reall, {\it 
A rotating black ring in five dimensions,}
Phys. Rev. Lett. {\bf 88}, 101101 (2002), hep-th/0110260.

\bm{ps} A.A. Pomeransky and R.A. Sen'kov, {\it Black ring with two angular
momenta,} hep-th/0612005.

\bm{lumeipope} H. L\"u, Jianwei Mei and C.N. Pope, {\it New black holes
in five dimensions}, arXiv:0804.1152 [hep-th].

\bm{plebdemi} J.F. Plebanski and M. Demianski,
{\it Rotating, charged, and uniformly accelerating mass in general
relativity},
Annals Phys.  {\bf 98} (1976) 98.

\bm{chenlupope1} W. Chen, H. L\"u and C.N. Pope,
{\it Kerr-de Sitter black holes with NUT charges}, 
Nucl. Phys. {\bf B762}, 38 (2007), hep-th/0601002.

\bm{chenlupope2} W. Chen, H. L\"u and C.N. Pope,
{\it General Kerr-NUT-AdS metrics in all dimensions}, 
Class. Quant. Grav. {\bf 23}, 5323 (2006), hep-th/0604125.

\bm{gilupapo1} G.W. Gibbons, H. L\"u, D.N. Page and C.N. Pope,
{\it The general Kerr-de Sitter metrics in all dimensions,}
J. Geom. Phys. {\bf 53}, 49 (2005), hep-th/0404008.

\bm{gilupapo2} G.W. Gibbons, H. L\"u, D.N. Page and C.N. Pope,
{\it Rotating black holes in higher dimensions with a cosmological
constant,} Phys. Rev. Lett. {\bf 93}, 171102 (2004), hep-th/0409155.

\bibitem{myper}
R.C. Myers and M.J. Perry, {\it Black holes in higher 
   dimensional space-times}, 
Annals Phys. {\bf 172}: 304 (1986).

\bm{kk} D. Kubiznak and P. Krtous,
{\it On conformal Killing-Yano tensors for Plebanski-Demianski family of
  solutions,} Phys. Rev.  {\bf D76}, 084036 (2007),
arXiv:0707.0409 [gr-qc].

\bm{cvetyoum} M. Cveti\v c and D. Youm,
{\it General rotating five dimensional black holes of toroidally 
compactified heterotic string},
Nucl. Phys. {\bf B476}, 118 (1996), hep-th/9603100.

\bm{cclp} Z.W. Chong, M. Cveti\v c, H. L\"u and C.N. Pope,
{\it General non-extremal rotating black holes in minimal five-dimensional
gauged supergravity,} Phys. Rev. Lett.  {\bf 95}, 161301 (2005)
hep-th/0506029.

\bm{emparan} R.~Emparan, {\it Rotating circular strings, and infinite
non-uniqueness of black rings,} JHEP {\bf 0403}, 064 (2004),
hep-th/0402149.

\bm{elvempfig} H. Elvang, R. Emparan and P. Figueras,
{\it Non-supersymmetric black rings as thermally excited supertubes},
JHEP {\bf 0502}, 031 (2005), hep-th/0412130.

\bm{mo} S. Mizoguchi and N. Ohta,
{\it More on the similarity between $D = 5$ simple
supergravity and M theory,} Phys. Lett.  {\bf B441}, 123 (1998),
hep-th/9807111.

\bm{cjlp} E. Cremmer, B. Julia, H. L\"u and C.N. Pope,
{\it Higher-dimensional origin of $D = 3$ coset symmetries,}
hep-th/9909099.

\bm{bccgsw} A. Bouchareb, G. Clement, C.M. Chen, D.V. Gal'tsov,
N.G. Scherbluk and T. Wolf, {\it $G_2$ generating technique for minimal
$D=5$ supergravity and black rings,} Phys. Rev. {\bf D76}, 104032 (2007),
arXiv:0708.2361 [hep-th].

\bm{cvlupapo} M. Cveti\v c, H. L\"u, D.N. Page and C.N. Pope,
{\it New Einstein-Sasaki spaces in five and higher dimensions},
Phys. Rev. Lett.  {\bf 95}, 071101 (2005), hep-th/0504225.

\bm{cophor} K. Copsey and G.T. Horowitz,
{\it The role of dipole charges in black hole thermodynamics}, 
Phys. Rev.  {\bf D73}, 024015 (2006), hep-th/0505278.



\end{thebibliography}
\end{document}